\def\H0{$H_0$ = 100 {\it h} km s$^{-1}$ Mpc$^{-1}$}

\def\mucentral{mag arcsec$^{-2}$}
\documentclass[12pt,preprint]{aastex}
\slugcomment{Accepted in The Astronomical Journal} %

\shortauthors{Galaz et al.}
\shortauthors{Bulge evolution in spirals and LSB galaxies}

\begin{document}
%\baselineskip 2.5m

%------------------------ TITLE -----------------------------

\title{Bulge evolution in face-on spirals and low surface brightness galaxies}

\author{Gaspar Galaz\altaffilmark{1}, Alvaro
Villalobos\altaffilmark{2}, Leopoldo Infante\altaffilmark{3}}
\affil{Departamento de Astronom\'{\i}a y Astrof\'{\i}sica, 
Pontificia Universidad Cat\'olica de Chile, Casilla 306, 
Santiago 22, Santiago, Chile}
\author{Carlos Donzelli\altaffilmark{4}}
\affil{Grupo de Investigaciones en Astronom\'{\i}a Te\'orica y Experimental
(IATE), Observatorio Astron\'omico, Universidad Nacional de C\'ordoba,
Laprida 854, X5000BGR, C\'ordoba, Argentina}
\altaffiltext{1}{Visiting Astronomer, Las Campanas Observatory,
ggalaz@astro.puc.cl.} 
\altaffiltext{2}{Present address: Kapteyn Astronomical Institute,
Postbus 800 9700 AV Groningen, Netherlands. villalobos@astro.rug.nl.}
\altaffiltext{3}{linfante@astro.puc.cl.}
\altaffiltext{4}{charly@oac.uncor.edu.}

\begin{abstract}

It is an observational fact that bulges of spiral galaxies contain a high fraction of 
old and metal-rich stars. Following this observational fact, 
we have investigated colors of 21 bulges hosted by
a selected sample of high surface brightness spirals and low surface
brightness galaxies observed in $B$ and $R$ optical bands and in $J$
and $K_s$ near-IR bands. 
Using structural parameters derived from these observations we
obtain evidence that bulges could be formed by pure
disk evolution (secular evolution), in agreement with
the suggestion by some authors. The color profiles, especially the
near-IR ones show null or almost null color gradients, supporting the
hypothesis that the disk stellar populations are similar to those
present in the bulge, and/or some bulges can be understood as disks with enhanced
stellar density (or pseudobulges). In the optical, half of the galaxies present
an inverse color gradient, giving additional evidence in favor of secular
evolution for the sample investigated. The comparison of the 
observed colors with those obtained from spectrophotometric  
models of galaxy evolution suggests that bulges of the selected sample
have solar and subsolar metallicity, and are independent of the current
stellar formation rate. Also, we obtain evidence suggesting that
galaxies hosting small bulges tend to be systematically metal poor
compared to those with larger bulges. These results are being checked
more carefully with high S/N spectroscopy. 

\end{abstract}


\section{Introduction}

Low surface brightness (LSB) galaxies have become important members of
our extragalactic zoo. Since \citet{disney1976} showed that the 
central surface brightness limit of $\mu_0(B) = 21.65$ \mucentral
~ in spirals claimed by \citet{freeman1970} was just a selection effect 
due to the narrow surface brightness sensitivity of photographic 
plates, the discovery of LSB galaxies in huge numbers in the
universe gave rise to a new field of study in astronomy: the low surface brightness 
universe. Since then, a wealth of observational data
\citep{longmore1982, bothun1985, bothun1991, mcgaugh1995, oneil2000a}
and theoretical work  \citep{mcleod1995, sprayberry1997, deblok1997, 
dalcanton1997a, mcgaugh1998a, chiueh2000, bell2002, mcgaugh2003, swaters2003} 
has been invested not only to describe the
most relevant features of these galaxies, but also to analyze whether 
these features are consistent with models of galaxy formation and 
evolution. In the observational front, LSB galaxies have been investigated in the 
optical \citep{dalcanton1997b, bergvall1999, burkholder2001} in the 
near-IR \citep{romanishin1982, bergvall1999, bell2000, galaz2002, monnier2003}, in radio
wavelengths, including studies of the HI distribution
\citep{hoeppe1994, longmore1982, deblok1996, matthews2001a}, 
detected and mapped in CO 
\citep{oneil2000b, matthews2001b}, and also studied spectroscopically 
\citep{impey2001, bergmann2003}, including the analysis of rotation
curves \citep{mcgaugh2001, deblok2001} and the Tully-Fisher 
relation \citep{sprayberry1995, chung2002}. 

These references 
strongly indicate that LSB galaxies are unevolved systems with
low stellar formation rate (SFR), low metallicity, small stellar density, 
relatively high gas fractions and, from dynamical information, large amounts of
dark matter.
%\footnote{Although cold dark matter (CDM) is the most popular explanation 
%of the rotation curve flatness observed
%in most LSB galaxies, several authors appeal to 
%Modified Newtonian Dynamics (MOND, Milgrom 1983), as a better model to reproduce such
%a flatness, without the necessity of dark matter of any kind. 
%See for instance \citet{mcgaugh1998b}.}. 
However, several challenging questions still remain. First, in spite of the
evidence indicating that 
LSB galaxies are young when compared with spectrophotometric models, the 
spread in color is wide, implying that they have probably gone through 
diverse evolutionary paths. Moreover, the fact that only small amounts of
blue stars are needed to make bluer the integrated colors, due to the 
small stellar density, makes some LSBs appear spuriously 
younger than they really are. Or, equivalently, small changes in the SFR
make galaxies look much younger what they actually are.
Second, while emission line analysis of a few spectra of LSB galaxies 
have revealed the metallicity of the current {\em gas phase} \citep{impey2001,
bergmann2003}, the metallicity of the {\em underlying 
stellar population}, as measured for example using absorption lines, has
not been possible to be determined, except in some few cases
\citep{bergmann2003}. Third, the total stellar mass of LSB galaxies is still
poorly estimated, mostly due to the large uncertainties in the stellar
mass-to-light ratios at optical wavelengths. Only recently, less scattered 
M/L ratios have been derived using near-IR bands, where the extinction due to 
dust has a smaller effect than in optical
wavelengths \citep{galaz2002}. Finally, recent modeling of LSB galaxies
predict that, under the assumption that blue LSB galaxies are currently undergoing
a period of enhanced stellar formation, {\em there should exist} a population
of {\em red}, non bursting, quiescent LSB galaxies \citep{gerritsen1999,
oneil2000a}. 

In the context of the old stellar content of disk and LSB galaxies,
the bulge or nuclear component plays a fundamental role as it also
affect any analysis where rotation curves are studied
\citep{valenzuela2003,rhee2003}. Although by the proper Hubble
morphological definition late type spirals tend to be bulgeless,
recent studies with the HST have shown that many late type, face-on
spirals apparently bulgeless, have in fact small, tiny bulge
\citep{boker2003} missed by studies where the 
spatial resolution was not good enough to resolve such small
bulges. One of the recurrent questions is whether many of these
bulges hosted by LSB galaxies or disks with low stellar density, are
in fact high surface brightness bulges embedded in LSB disks. 
In such a case the bulge could evolve independently
with respect to the disk, and one should expect that bulges in LSB
disks present more or less the same metallicity than bulges of HSB
galaxies. 

On the other hand, the secular evolution of galaxies, i.e. the idea
that the bulge is formed from disk evolution, allowing the direct
formation of bulges from disks in isolated galaxies, implies that the
stellar population of disks are similar to those in bulges. For
example, one can expect that color gradients are less violent, or even
inverse (red to blue, from the center to the outskirt of the
galaxy). Also, one can expect that light profiles are fitted only by a
combination of exponential profiles, instead of the classic de
Vaucouleurs + exponential profile \citep{courteau1996,
kormendy2005}. In this sense, secular evolution could produce what is
called a pseudobulge, structurally similar to a bulge, although more
comparable to disk when its stellar population is analyzed and/or its
light and color profile is studied.  

In this paper we explore possible evolutionary scenarios for
a set of selected face-on spirals and LSB galaxies, 
using a combination of optical and near-IR colors. In particular, we
investigate the observables and structural parameters in scope of the 
secular disk to bulge evolution, as stated for example in \citet{courteau1996}.  
Most galaxies investigated here are {face-on \em nucleated} LSB galaxies; all
of them are part of the catalogue published by \citet{impey1996}. We
also investigate the stellar content of the bulges. The goal 
of this paper is to determine what ranges of metallicities and ages are
consistent with bulge and disk formation scenarios, by comparing 
observed $B$, $R$, $J$ and $K_s$ magnitudes and colors with
spectrophotometric models of galaxy evolution. We
have selected a subsample with prominent bulges, for 
which we have carefully studied their structural parameters (bulge and
disk scale lengths, surface brightnesses). For this subsample, we
present radial brightness profiles and color gradients, both in
optical and near-IR bands. Secular evolution is analyzed in scope of
these results. In \S 2 the sample
selection is presented. Observations and data reduction are described
in \S 3. In \S 4 we present results and a basic analysis. In \S 5 we
present a discussion in scope of the bulge formation in HSB and
LSB galaxies. We conclude in \S 6.

\section{Sample selection and previous observations}

Galaxies presented here were originally selected from the catalogue of
\citet{impey1996}. The sample includes galaxies with $21.0 \le
\mu_0(B) \le 23.5$ \mucentral, with HI masses such that
$7.49 < \log(M_{HI}/M_\odot) < 10.69$, and
with DEC $< 5$ deg, covering all RA.  
Our sample includes not only LSB galaxies, but HSB galaxies
as well. This allows to generate a continuous sequence for properties derived
for LSB galaxies and HSB galaxies, like colors, structural parameters, 
etc. \citep{galaz2002}. The total original catalogue amounts 107 galaxies which
satisfy these constraints.  77 galaxies were already observed in the
near-IR at Las Campanas Observatory between 1999 and
2000. Observations and reductions for this data set are described in
\citet{galaz2002}. From these 77 galaxies with near-IR photometry, 55
galaxies were observed during photometric nights in $B$ (Johnson) and $R$
(Cousins). In this paper, we focus our attention on a subsample of 21
face-on nucleated galaxies out of these 55 spirals. In order
to prevent undesirable strong internal reddening 
due to the effect of inclination, we have selected only face-on
nucleated spirals (defined following
Hubble morphological classification scheme) for which we were interested
to reach faint isophotes to $\mu_0(B) \sim 25.0$ \mucentral. The face-on
criteria is defined such that galaxies are spirals with inclinations 
smaller than 20 deg\footnote{Note that galaxy 515 is
probably at the limit of this criterion, having an inclination angle
of $i \sim 45$ deg. This turns its disk surface 
brightness about $2.5\log[\cos (i)] = 0.37$ magnitudes fainter, at most.}.  
In fact, the combined optical and near-IR imaging \citep{galaz2002} 
allow us to select galaxies with prominent bulges. The 21 spirals
exhibit central surface brightnesses between 21 and 23 \mucentral, in
the $B$ band, allowing to reach the desirable central surface
brightness limit. Figure \ref{mosaics} shows an optical mosaic of the
21 galaxies selected for the main part of this study. Note that all
galaxies present bulges. Also note that 6 galaxies are barred. 

\placefigure{mosaics}

\section{Observations and data reduction}

\subsection{Observations}

Optical observations were strictly performed during photometric
nights in three runs at Las Campanas Observatory, using the 2.5m du Pont 
telescope. The first two runs took place in April and August 2000, and the 
third one was in April 2002. In all of them we used the same detector: a 
2048$\times$2048 Tek\#5 CCD of 0.259 arcsec/pixel 
scale and 8.8 arcmin FOV. The CCD present low read out noise (4.2 e$^-$)
at a gain of 3.0 e$^-$/ADU. Typical exposures were between 800 and 
1200 secs, splitted in two exposures per galaxy per filter. Typical
seeing was around 0.9 and 1.2 arcsec in the $B$ band, and 0.8 and 1.0
arcsec in the $R$ band. 

\subsection{Reductions}

With the purpose of obtaining an acceptable accurate photometry for our
sample, it is important to reduce very carefully our set of
images. Special attention was taken in the flat-fielding  
process and fringing corrections, particularly in the $R$ band. 
In fact, low (flat) and high (fringes) frequency spatial variations can
introduce a spurious systematic signal over the CCD that affects
the surface brightness estimate of our galaxies. In order
to reduce as possible these kind of effects, we build-up a super flat for each
filter from observations, especially in $R$, where fringes
dominate the spatial signal variations at high frequencies. A 
Fourier transform of the images allows to detect the most
common spatial patterns. Also, and before 
flat-fielding, images were cleaned-up from cosmic rays, bad pixels and
dead columns. Images were finally aligned\footnote{During observation
for each galaxy, telescope was shifted 3 arcsec ($\sim 10$ pixels)
between consecutive exposures to allow for photons of the same source
to fall into different pixels.} and averaged.  

\subsection{Photometric calibrations}

After removing instrumental effects and pixel to pixel variations, 
images were photometrically calibrated. The usual
method based on observation of Landolt standards \citep{landolt1992}
was used.  
We observed no less than 15 standards per night, allowing to 
derive accurate zero points and color terms for our transformation 
equations
\begin{eqnarray}
B & = & m_B + Z_B + k_B * X_B + C_B * (m_B - m_R)  \\ 
R & = & m_R + Z_R + k_R * X_R + C_R * (m_B - m_R),
\end{eqnarray}
where $B$ and $R$ are the calibrated magnitudes in the standard 
system; $m_B$ and $m_R$ are the instrumental magnitudes; $Z_B$ and 
$Z_R$ are the photometric zero points; $k_B$ and $k_R$ are the
atmospheric extinction coefficients; $X_B$ and $X_R$ are the measured 
airmasses at the moment of the observation; and $C_B$ and 
$C_R$ are the first order color terms. All photometric reductions were
performed with the IRAF PHOTCAL package. 

Solving the two above equations independently for each night it is 
possible to determine the photometric quality of each night. 
The corresponding coefficients for each run are presented in 
Table \ref{phot_coefs}. Errors of aperture photometry
were computed using the standard rules of error propagation from
original measured fluxes, taking into account the S/N ratio of the
different stars and galaxies and also the detector features (read out
noise and gain). In the error propagation we also take into account
the effect of the various steps involved within the image reduction
process (bias subtraction, flat field division, summing, etc...).  The
calibration for the corresponding $J$ and $K_s$ magnitude for each
galaxy was  already presented in \citet{galaz2002}.

Table \ref{magnitudes} present $B$ and $R$ magnitudes for all spirals
in the sample (from which a subsample of 21 face-on nucleated galaxies is
taken, denoted as bold-face identifications). Magnitudes are
calculated for the inner 2 kpc diameter, which is 
estimated using the radial velocity of each galaxy and the diameter
angular relation with a Hubble constant H$_0$ of 75 km sec$^{-1}$
Mpc$^{-1}$. Considering the typical size of the
galaxies (see next section), we estimate that inside the inner 2 kpc
radius we are including the typical bulge population. 
Along with the atmospheric extinction correction, we have applied to
all images Galactic extinction corrections using the maps of
\citet{schlegel1998}. 

\subsection{Computing light profiles, color
profiles and structural parameters} 

Structural parameters are consistently computed using the light
profiles in the four bands $B$, $R$, $J$ and $K_s$. These are derived 
using ELLIPSE (IRAF), which provide
azimuthally averaged isophotes for each galaxy. From these values one
can derive fitting parameters for both the bulge and  disk components
simultaneously \citep{beijer1999}. From the isophote 
level we derive exponential profiles of the form 
\begin{equation}
\Sigma(r) = \Sigma(0) \exp^{-(r/h)},
\label{exp_profile}
\end{equation}
where the structural parameters defining these profiles are 
$\Sigma(r=0)$, the surface brightness at $r=0$, in L$_\odot$ pc$^{-2}$; 
and $h$, the scale length of the bulge or disk, i.e., the radius at
which we have $1/e$ times the central intensity. Transforming equation
\ref{exp_profile} to a logarithmic scale we have
\begin{equation}
\mu(r) = \mu_0 + 1.086 (r/h),
\label{exp_prof_log}
\end{equation}
where $\mu_0$ is the central surface brightness in \mucentral~ for the 
disk, and $r$ is the radial distance to the azimuthally averaged
isophote.  
For the bulge component it is also common to use the 
$r^{1/4}$ profile law in the form
\begin{equation}
\mu(r) = \mu_e + 8.325\left[(r/r_e)^{1/4} - 1\right],
\label{old_profile}
\end{equation}
to fit the bulges of all kind of spiral galaxies (as they were similar
to elliptical galaxy profiles), where $\mu_e$ is the surface brightness
at $r_e$, the so-called effective radius containing half of the galaxian
light. In fact, we do know that the exponent of the fit is correlated
with the morphological type: light profiles of ellipticals and S0s 
follow the $r^{1/4}$ law \citep{devaucouleurs1948}, and bulges 
of Sa and Sb fit better to $r^{1/2}$ profiles \citep{beijer1999}. 
%More generaly, a S\'ersic
%profile is in general the best choice, with a general exponential profile
%$(r/r_e)^{1/n}$ \citep{sersic1968}. 

We use the task NFIT1D in STSDAS/IRAF to fit two exponential (and $r^{1/4}$)
profiles for the whole light curve, using a $\chi^2$
minimization algorithm. Light profiles errors are related 
to the corresponding errors in the fitting
parameters. These errors are derived from usual $\chi^2$
minimization. In turn, each isophote has an associated error given by
the ratio between the root-mean-square scatter around the isophotal
intensity and the square root of the total pixels included in the
isophote. 

Table \ref{structural_parameters} shows optical structural parameters measured
for the sample of 21 spirals. Table \ref{structural_parameters_ir}
shows the corresponding parameters measured from the near-IR brightness
profiles. Parameters include central surface
brightnesses and scale lengths for both the bulge and the disk in the
four bands. Note that 38\% of our sample is composed by bonna fide LSB
spirals (fainter than 22.0 \mucentral), and the rest are in fact HSB spirals.  
This is useful to clearly determine what properties in spirals suffer a 
continuous transition to the LSB regime and which do not. In particular, we are 
interested on colors and light profiles for our sample, since its
behavior is one of the diagnosis for a bulge/disk
evolution (see \S 5). 

In this work, all disks are fitted better by an exponential profile
of the form given by equation \ref{exp_profile}, or equivalently by its
logarithmic version given by equation \ref{exp_prof_log},
from where the central surface brightness of the disk 
is computed. For bulges, the usual method is to use a $r^{1/4}$
profile (or de Vaucouleurs profile). However, we shall demonstrate
that most of our galaxies are better fitted by a
combination of {\em two} exponentials: one for the disk and one for
the bulge. 
Figure \ref{profiles} shows the surface brightness profiles and color
gradients in $B$ and $R$ for the 21 galaxies selected for the bulge
analysis. Figure \ref{near_ir_profiles} present the corresponding
near-IR surface brightness profiles and color gradients.
Names are correlative to the \citet{impey1996} catalogue.
Dashed lines represent exponential fits, using equation
\ref{exp_prof_log}. In Figure \ref{profiles}, we 
include the fitted parameters $\mu_0$ in $B$ and $R$, expressed in
\mucentral, and $h$, the scale length, expressed in kpc. Vertical
lines in the panels of color gradients, represent approximated
boundaries for {\em pure} bulge and disk components, computed
visually. Light and color profiles extend up to the radius where the
1$\sigma$ isophote is reached. For 16 out of 21 galaxies the bulge
component can be fitted in the 
optical by an exponential profile (equation \ref{exp_profile}) instead
than a de Vaucouleurs  profile (equation \ref{old_profile}). In the
near-IR, 14 out of 21 galaxies, are well fitted by two
exponentials. The radius separating the bulge and the disk components
is estimated visually, and then computed using a  simultaneous
exponential fit for the bulge and the disk. Note that 5 galaxies
which are better fitted in the optical by exponential profiles are
however better fitted by $r^{1/4}$ profiles in the near-IR. Also note
that some galaxies with good fits in the near-IR are not fitted in the
optical neither by an exponential neither an $r^{1/4}$ profile (see
below). 

In most cases, a fit using the $r^{1/4}$ law was unsatisfactory for 
the central part of the galaxies.  
This functional form gives overestimates the bulge central brightness
$\mu_0$(bulge), as has been noted by \citet{beijer1999} and \citet{galaz2002}.
The exponential fit gives better results for our sample of galaxies.
Five galaxies present no definite optical detection for the disk, and
therefore no fitting was possible. For galaxy 473, no acceptable
fit was possible to obtain in these bands. For galaxy 264 we were
unable to derive a good fit in $J$ and $K_s$ for the bulge
component.    
It is also possible to appreciate for some cases the ``flattening'' effect in the 
central zone of the fittings, due to the seeing effect. This effect,
however, is so small that it does not change our results\footnote{We
note that our worst seeing was about 0.9 arcsec, which, considering the average
distance of our galaxies, does no affect the brightness nor the color
radial profiles.}.

\placefigure{profiles}

\placefigure{near_ir_profiles}

\section{Results and analysis}

In this section we present and discuss the color-magnitude and the 
color-color diagrams. We also use some measured structural parameters
which help to study stellar populations in the bulge of our
sample galaxies, and analyze possible secular evolution of the
galaxies. We also discuss light and color profiles in optical and
near-IR bands. We compare our results with those obtained for
HSB galaxies by other authors. 
%Finally, we use the spectrophotometric
%model of galaxy evolution (for different metallicities) given 
%by Bruzual \& Charlot (1996) to constrain age and 
%metallicity of galaxy nuclei. 

\subsection{Morphology and color-magnitude diagram}

Morphological classification is important because it links those
features observed locally with those observed at high-$z$. 
\citet{galaz2002} showed that bluer, gas rich galaxies are more
irregular compared to those with lower fractions of HI. Figure
\ref{color_mag_BR} show that the 21 face-on nucleated galaxies selected are in
fact brighter and generally redder than most of the rest of the sample. 
%Our original sample of 55 galaxies observed with $B$ and $R$ 
%filters show that bluer, gas rich galaxies are more irregular and
%generally bluer (in Figure \ref{color_mag_BR}, ``other'' means
%Irregular and dIn types), as
%has been also shown in \citet{galaz2002}. 
It is also true that, concerning light emitted by stars, bluest
galaxies have the lowest absolute magnitudes (Figure
\ref{color_mag_BR}),  a trend largely observed in the universe and the 
first indication that (1) most of the light
emitted by very blue galaxies does not come from their continuum
stellar emission but from the gas emission, and (2) they present
generally an irregular morphology.
We note that we have indirectly excluded in our sample dwarf
spheroidals (dSph) and dwarf ellipticals (dE). Although these galaxies
(especially dSph) present in general low surface brightnesses and red
colors,  these are not present in our sample, since ours is selected
from the Impey et al. (1996) catalog, which lack these types of
galaxies. We recall that the Impey catalogue of spirals was built
mainly from the APM catalogue \citep{loveday1996}, which in fact lack
dSph and other kind of red but faint galaxies. However, we have
checked that the luminosity of bulges hosted by spirals is wide, and
the sample include
all the Hubble classification for normal spirals (from Sa to late Sc). 
%The lack of {\em red} galaxies at faint absolute
%magnitudes in Figure \ref{color_mag_BR} represents the absence of dwarf 
%ellipticals and dwarf spheroidals in our sample, and in fact 
%in the original Impey catalogue. 

\placefigure{color_mag_BR}

\subsection{Structural parameters}

Here we discuss some structural parameters for the 21
face-on nucleated spirals (Figure \ref{mosaics}). Brightness profiles
and color gradients  are presented in Figure \ref{profiles} (optical
bands) and Figure \ref{near_ir_profiles} (near-IR). 

Figures \ref{mu_h} and \ref{mu_mu} show that our sample
follows the same trends observed by the galaxies of \citet{dejong1995},
showing that we have not selected spirals with very bright bulge surface 
brightness (note the outlier LSB 264 with $h_d = 0.34$, with a large
central region compared to the disk, see Figure 1). From Figure \ref{mu_mu} it is clear that
galaxies with brighter bulges have brighter disk surface
brightness. The range of bulge surface 
brightness is consistent with the fact that the original sample from
the Impey et al. (1996) catalogue (which selects galaxies from the APM survey)
includes the total range of spiral galaxies, from Sa to late Sc. We
note that our sample of 21 galaxies is in any case a complete
sample: these are simply the number of galaxies from our sample with
valid $B$ and $R$ photometry which are oriented face-on.  

On the other hand, Figure \ref{BR_mu} shows the inner 2 kpc $B - R$
color as a  function of the disk central surface brightness. This Figure shows
that there is no bias or selection effect between color and surface
brightness: a wide color range is included in the sample for both the 
faint and bright end. However, note that there are tight correlations
between the HI gas mass (taken from
\citet{impey1996}), the color, the absolute total magnitude and the
surface brightness. Figure \ref{relations} shows the relationship for
these parameters, for the 52 galaxies shown in Table
\ref{magnitudes}. The 21 nucleated, face-on galaxies (black dots) are
in fact part of the most massive sample (in terms of the HI content),
and with the brightest absolute magnitudes. The color $B-R$ is computed for the inner
2 kpc for each galaxy. Note the tight relationship for the parameters,
especially for the absolute magnitudes in both filters $B$ and
$R$. The brighter a galaxy is, the larger the HI mass, the redder the
galaxy, and the higher the surface brightness are. These trends in
optical bands agree with results of \citet{galaz2002} in the near-IR
bands $J$ and $K_s$.  However, note that one can infer from Figure
\ref{relations} that LSBGs ($\mu(0) > 21.6$ \mucentral) have faint
absolute magnitudes ($M > -16$ mag). This is different to what was
found in the near-IR bands \citep{galaz2002}.   

\placefigure{mu_h}
\placefigure{mu_mu}
\placefigure{BR_mu}
\placefigure{relations}

\subsection{Color gradients}

Figure \ref{profiles} presents the surface brightness and color
profiles for the 21 face-on nucleated galaxies. Contrary to what was
expected, a high fraction of them (10 galaxies out of 21)
present a significant inverse color gradient near their center: the
color is redder as the distance to the bulge increases. Could this due
to dust present in the disk, which redden the color? If this is not
the case, could this be a sign of bulge secular evolution in these
spirals, i.e., the bulge was formed after the disk? 

Currently, many authors argue that the dust content of the disk in LSB
galaxies is low \citep{deblok1995}, and there is some evidence
suggesting that some LSB galaxies are 
dust poor \citep{beijer1999}. The combined near-IR and optical photometry
shows that these galaxies in fact do not have significant
amounts of dust in their most inner parts and their bulge
neighborhood (see however \citet{holwerda2005}), and is unable to
reproduce the observed color gradients 
(e.g. galaxies 196, 213, 345, 463, 468, 470, 471, 473, 484 and 488, see
also discussion in \S 5.1). 
%The
%average gradient between the inner and outside part of the bulges is
%about $\sim 0.32$ mag, well beyond the photometric error and the
%amount of extinction expected from the $B - R$ and $J - K_s$ measured
%differences, up to the inner $\sim 20$ kpc. Only one galaxy has a null
%color gradient (galaxy 207), curiously a ringed galaxy.  

However, near-IR color gradients (Figure \ref{near_ir_profiles}) show
that most of the near-IR surface brightness profiles are (1) well
fitted for all their extension solely by the combination two
exponential profiles, and (2) color gradients are null. 
Although this last point could support by itself the hypothesis that
reddening is not strong in these galaxies or, at least, the dust
abundance is similar along the galaxy components (bulge or disk), a
large amount of dust seems not to change necessarily the near-IR
colors, and therefore the color gradient, as shown in \S 5.1. 

% ......... INSERT FIGURE WITH THE NEAR-IR COLOR GRADIENTS OR AT LEAST
% SOME VALUES WHICH CAN SUPPORT OR RULE OUT THE PRESENCE OF DUST.  

Therefore it is likely that for our selected sample
of spirals and LSB galaxies, the nature of these color gradients is
not the  dust content, either for the galaxies presenting ``normal''
color gradients (i.e. galaxies with nearly flat or red to blue color
gradients), as well as for those presenting inverse color 
gradients. 

For galaxies with normal color gradients, we support the hypothesis
that observed color gradients are generated by a canonical stellar
distribution along the radial axis: the stellar 
populations more distant to the bulge are metal poorer and/or younger 
compared to the stellar population closer to the bulge. This
scenario agree with the idea that the bulge is formed first from
accretion, and the disk form after, and does not agree with stellar
formation occurring  progressively from the disk to the 
center of the galaxy mass distribution (the bulge secular evolution). 
At first impression, the bulge secular evolution does
not match with the observation that some early type galaxies were
already formed at high-$z$ ($z \sim 2$), following recent results from
the Hubble Ultra Deep Field (HUDF). However, there could be an
undetermined fraction of bulges in the HUDF hosted by undetectable disks
with central surface brightnesses well below the limiting surface
brightness reached by the ACS at the HST\footnote{This
surface magnitude ``drop out'' selection for high-$z$ galaxies is
mainly due to the cosmological dimming factor $(1+z)^{-4}$ for the
surface brightness \citep{tolman1930}.}. These spiral galaxies could
be therefore misclassified as pure ellipticals (compact
ones). In spite of this evidence, we wish to confront the model of
bulge secular evolution with our analysis from the structural
parameters.  

Assuming that the internal dust extinction is not the main reason for the
inverse color gradients for the 10 galaxies of our sample,
the hypothesis of secular evolution for these galaxies should
be seriously considered. We investigate and discuss this hypothesis in
the next section, especially in scope of the results for the near-IR
light profiles and color gradients.

\section{Discussion}

\subsection{Application of the spectrophotometric model in the context
of bulge and disk formation}

Colors observed in cores of spiral galaxies selected for
this study can be compared to those predicted by
spectrophotometric models of 
galaxy evolution. These models predict, among many other quantities, the 
integrated colors and ages for stellar populations, for a given range
of metallicity. 

The synthesis models use a wealth of information about stellar
evolution, which include stellar evolutionary sequences, spectral
libraries, and a number of tunable parameters such as the initial
mass function (IMF) slope, the stellar formation 
rate (SFR), and current results of semi-empirical theories of chemical
enrichment. They allow the computation of spectral energy distributions
for an ensemble of stars (a galaxy), and some selected spectral
indices and colors, as a function of age and metallicity.
The model used here is that of \citet{bruzual2003}, which
provides results as a function of the mean age, after an 
instantaneous burst of stellar
formation and subsequent exponential (increasing and decreasing) SFRs, for
a metallicity range 0.005--2.5 Z$_\odot$. The IMF used is that of 
\citet{scalo1986}. The \citet{salpeter1955} IMF gives similar results. 
Different amplitudes for the star formation are indicated by different
exponents in the SFR, which can increase or decrease exponentially
from an initial time $t_0$ to present, and $\tau > 0$ for SFRs 
$\propto e^{-t/\tau}$, and $\tau < 0$ or SFRs $\propto e^{(t_0-t)/\tau}$. 
Each one of these exponents is associated with a mean stellar age for the
corresponding stellar population. Giving the joint evidence of WMAP
that the last electron scatter was about 13.7 billion years ago, and
stars formed 200 million years after \citep{bennet2003}, it is
reasonable to fix model star formation 12 Gyrs ago. The characteristic
time $\tau$ is related with the stellar formation efficiency. For example,
the most efficient SFR is set with $\tau = -1$, where most of the population
is formed recently, and has 1 Gyr mean age. Alternatively, for the less 
efficient stellar formation we use $\tau = 0$, corresponding to an
instantaneous burst, i.e., all stars formed at the same time, with no
further stellar formation processes. 
%The mean age of this kind of population is 12 Gyrs. 
%
%\subsection{Spectral model and bulge formation in HSB and LSB
%spirals}
%

As discussed in the next section, current evidence suggests that
possible paths 
of bulge formation in HSB galaxies are closely related to the relative 
size of bulges \citep{courteau1996}. Bulges of HSB spirals, large and
prominent, share some 
properties in their stellar populations with those featured by medium
size elliptical galaxies. More precisely, both populations of stars
exhibit the same age and metallicity (old and metal rich). 
In contrast, {\em small} bulges present in a number of HSB spirals clearly
exhibit different abundances and probably have different ages 
compared with medium size ellipticals, {\em but} similar properties
compared with stellar populations in {\em disks} (younger and less
metal rich than core of ellipticals). Are these features also observed 
in LSB galaxies? 

The color-color diagram $J - K_s$ vs. $B - R$ for the inner part of
the bulges (radius $< 2$ kpc) 
and the corresponding color-color model grid are shown in 
Figure \ref{BR_JK_2}. Black dots represent bulge colors of our 21
nucleated, face-on spirals and open circles of \citet{peletier1996}
data, corresponding to a $B$, $R$, $J$, and $K_s$ photometry for a sample of
bulges hosted by spirals. The major change in $J - K_s$ is for a
Z$_\odot$ population, which varies only $\sim 0.15$ mags for an
increase of mean age from 1 Gyr to 12 Gyrs. All points have been
corrected from atmospheric and Galactic extinction, this last using
the \citet{schlegel1998} extinction maps. 

\placefigure{BR_JK_2}

Figure \ref{BR_JK_1} shows the color-color diagram using colors of
the {\em inner 2 kpc} for the 21 selected galaxies, superimposed over
the \citet{bruzual2003} model grid as in Figure \ref{BR_JK_2}. Colors
have been corrected using the Galactic extinction maps of
\citet{schlegel1998}, and internal extinction supposing a maximum in
$B-R$ reddening of 0.3 mag (see next paragraph). This color-color
diagram suggests that stellar populations of small bulges (open symbols),
defined by the bulge to disk scale length ratio 
$h_b/h_d < 1\sigma$ (see next section) compared to average,
appear slightly younger and clearly metal poorer
than stellar populations hosted by larger bulges ($h_b/h_d >
1\sigma$, filled circles), except for
three bulges with metallicity above Z$_\odot$. This is expected from the
hypothesis of secular evolution: bigger bulges are result of larger
accretion time, generating more stellar formation in the past and
then the present stellar population is more metal rich compared to
what is found in smaller bulges. From Figure \ref{BR_JK_1} it is also
apparent that LSBGs\footnote{Note that here the term
HSB galaxy is defined for galaxies with $\mu(0,B) < 22.0$ \mucentral,
as usual, but note however that all galaxies are just above or tiny
below this limit, and hence we really do not have HSB galaxies
strictly speaking.} tend to have smaller bulges (open
squares), which is also consistent with secular evolution: a smaller
stellar stellar density (which define a LSBG) could impact on a slower
secular evolution, forming a smaller bulge as see today. In
summary, our data suggest that metal poor bulges tend to be smaller
than metal rich bulges, on average.   

\placefigure{BR_JK_1}

This result {\em suggests} that the 
relation between the relative size of HSB spiral bulges and their
eventual secular evolution, could be also applicable to LSB spirals,
given the melting  HSB/LSB fraction in our sample. 
%We have verified that near-IR magnitudes of
%the galaxy outside the grid was obtained during non-photometric conditions. 
From Figure \ref{BR_JK_1}, it is clear that 
$J - K_s$ is a good metallicity indicator \citep{galaz2002}, and
relatively insensitive to stellar populations of different mean
ages.

%
%\subsection{Internal extinction}

A very hard problem to solve is the amount of internal extinction
which reddens each particular galaxy. The optical color
$B - R$ is the most affected by dust and/or gas. In principle, the amount
of extinction could be 
very different from one galaxy to another. However,
it is important to keep in mind that in our case: (1) colors are
computed for the central part of the galaxies, in particular for the
bulges, where dust is not generally present; therefore it is
reasonable to assume that extinction should not be too large in these
regions. (2) The selected galaxies are oriented
face-on, and then the line of sight is not passing through a thick portion
of the corresponding galactic disk and/or spiral arms; thus, the extinction
should not be large. However,
when we observe the color-color diagram at face value, specifically
the $B - R$ color, one has the impression that $B - R$ is too red
compared to the models, 
appealing for a possible significant extinction. The key is to
estimate the modulus of the extinction vector in the color-color
diagram using some reasonable assumptions. For this we use the
spectrophotometric model PEGASE \citep{fioc1997}, which offer the
possibility to consider extinction, by allowing to generate a 
set of synthetic spectra with different amounts of dust, which are
convolved with synthetic filters to estimate the reddening vectors in
the $J - K_s$/$B - R$ plane.  
Results show that increasing the dust content of the bulge component
by as much as 60\% from a nominal value, imply an increase in $J -
K_s$ and $B - R$ colors by no more than 0.15 mag over the full range
of colors (see Figure \ref{pegase_reddening}). Each symbol in Figure
\ref{pegase_reddening} represent a different evolutionary stage (from
0.5 Gyr to 13 Gyr, from left to right) with a different metallicity,
consistent with stellar evolution, nucleosynthesis and supernova
rate. For each point or square, the dust {\em surface
density} is 60\% larger than a fiducial dust surface density at the
same age.

\placefigure{pegase_reddening}

In summary, we conclude that the internal extinction
should not be large, {\em at least} in the regions where our
color indices were calculated, i.e. in the bulges. This is discussed by
other authors: for example \citet{bell2000}, precisely claim that dust
does not have a large impact on the color of {\em face-on} nucleated LSB
spirals.  

%However, when we compare our observed $B - R$ colors with those from
%models, we have the impression that many bulges have redder colors 
%than expected, probably due to internal reddening. In fact, some 
%galaxies appear to be older than the limit age of 12 Gyrs impossed by
%the model. Note that older ages appear even closer to the 12 Gyr isochrone
%compared to the color difference between the 11 Gyr isochrone and the
%12 Gyr one. Recall that all magnitudes and colors have been corrected from
%Galactic extinction. 
An almost negligible extinction in $J - K_s$ and small extinction
in $B - R$ in the bulge (a maximum of $\sim 0.3$ mag) would primarily
affect our age estimates but not our metallicity determinations. Note
in Figure \ref{pegase_reddening} that there is an interval of age
where the index $B - R$ has a {\em lower} reddening than $J - K_s$,
probably due to a saturation effect on the $B - R$ index, which seems to
continue for $B - R \ga 1.2$, reflected in the decease of the color $B
- R$; while the
optical thickness increases, the $B - R$ 
color may saturate and the bulge gets optically thick in both
bands. At the same time, however, the galaxy  is optically thin in $J$
and $K_s$, although the opacity $\tau$ increases and $J - K_s$ still
reddens. 

Bulges of HSB galaxies appear to have less scattered colors than those
observed in spirals and LSB galaxies of our sample, as shown in
the color-color 
diagram. Their metallicity range is 1.0Z$_\odot <$ Z $< 2.5$Z$_\odot$
and ages are around 8.6 Gyr. One can compare these results with those
obtained for high surface brightness (HSB) bulges. For example,  
the higher metallicity measured for bulges of HSB spirals of
\citet{peletier1996}, represented as open circles in Figure
\ref{BR_JK_2}, is in agreement with the fact that these contain
evolved, enriched old stellar populations. However, our data disagree
with the metallicity measured for the HSBs in Peletier et al., in
the sense that they claim solar or near solar metallicities using 
the models of \citet{vazdekis1996}, and the model we use gives super-solar
metallicities. What probably makes the difference
between Vazdekis et al. models and the Bruzual \& Charlot models used
here is the contribution of the post-AGB stars and other advanced stages of 
stellar evolution. 

In the case of our normal and LSB spirals, we obtain that bulges show
solar or subsolar metallicities, and ages as 
young as 4 Gyrs. This supports well known results where LSBs appear
as metal poor systems \citep{deblok1998, bell2000}. We also measure
systematic bluer color for LSB bulges, as compared to those bulges of HSB
galaxies (see Figure \ref{BR_JK_2}), both in $B - R$ and $J -
K_s$. 
%We stress that very red (in $B - R$) bulges of LSB galaxies of our sample 
%are due most likely to age than metallicity. 
%This agrees with results
%of \citet{padoan1997}, who challenge the apparent young age of stellar
%populations of LSB galaxies from blue optical colors, consistent with
%those observed in old stellar populations which have very low
%metallicities (Z$\sim$0.0002Z$_\odot$).   

%Finally, we have some galaxies which present inverse color gradients
%(LSBs 213, 242, 345, 468 and 484). 

\subsection{Bulge formation scenarios for HSB and LSB galaxies}

One of the fundamental questions about galaxy formation, and in scope
of the observed stellar population content at a given redshift, is
the possible {\em chronological} order in which the different
structural components of a spiral galaxy form. This chronology clearly
imposes constraints to the galaxy formation models and to the
subsequent stellar evolution of the different building blocks of
galaxies. 
%The importance
%stands since this chronology imposes
%constraints on theory of galaxy formation and subsequent evolution,
%not only on the galaxy formation process as a whole, but also on the
%stellar evolution of the different building blocks of galaxies
%(gas, stars, bulge, disk, spiral arms, bars, dust, etc..). 
In this context, one can ask about the order in which the two major
components of spirals, namely the bulge and the disk, are formed.  

Many studies have pointed out that 
the bulges of grand design spirals (like M31) share many properties 
with those observed in medium size elliptical galaxies. 
Among these properties we can mention the so-called $D_n-\sigma$ \citep{dressler1987}
and the fundamental plane \citep{bender1992} relations. 
Furthermore, \citet{jablonka1996}
found that bulges of late type spirals (Sc) follow the same Mg-$\sigma$ 
relationship defined for early type ones, suggesting that their bulges share the 
same mass-metallicity relationship as observed in elliptical 
galaxies. These results, along with the kinematical evidence provided
by velocity dispersions, support the 
idea that medium-size elliptical galaxies accrete gas from the
surrounding environment, forming a disk and then building up a spiral
galaxy \citep{kauffmann1996}.  
In this way, stars in bulges of spiral galaxies should be similar 
to those in elliptical galaxies, i.e., old and metal rich, differentiating
from those populating disks.

Results from kinematical studies and surface brightness observations
of HSB galaxies with small bulges, indicate that they share many
properties with their corresponding disks.  
For example, many bulges of spiral galaxies are much 
better fitted by exponential
profiles than by de Vaucouleurs profiles \citep{dejong1996}, and also
many of them present a disk like kinematics \citep{kormendy1993}.
These results suggest that small bulges of spiral galaxies were built 
from stars already formed in the corresponding disks, which in turn
indicate that stars currently populating such bulges should be similar
to those stars populating the corresponding disks (relatively young
and metal rich, but less than those populating  
elliptical galaxies), and therefore are different to stars present in
elliptical galaxies.   
In fact, the excellent work of \citet{trager1998} based on stellar
population models and spectroscopy of the central region of Sa-Sd spirals,
provides evidence in favor of bulge formation in late spirals 
triggered by gravitational interaction in their disks. One can argue
that the above evidence supports secular evolution. 
Are these results also valid for some LSB spiral galaxies?

The hypothesis of secular evolution (that disk form first, and the bulge emerges
naturally from this \citep{courteau1996}), imply that the disk and
bulge scale lengths are correlated or, in other words, the
bulge-to-disk (B/D) ratio is independent of galaxy type. If the sequence
of galaxy morphology is continuous in terms of surface brightness, naturally 
the statement should be valid for both LSBs and HSBs. 
This correlation is supported by some
models \citep{combes1990, struck1991}, and depends mainly on the relative
timescales of the star formation and the viscous transport, and on
the strength of the total angular momentum. Also, one can predict that
secular evolution will produce in many cases what is called
pseudobulges: disks agglomerate to the center of the galaxy,
producing a high density of stars which structurally are similar to
bulges, and share similar stellar population properties observed in the
disk. One of the observed properties is that the entire surface
brightness profile can be fitted by a combination of two exponential
profiles. This is exactly what we obtain for most of the optical profiles
(14 galaxies, Figure \ref{profiles}) 
and almost all the near-IR profiles (16 galaxies) as presented in
Figure \ref{near_ir_profiles}. Thus, one can argue that our data
support the evidence not only of secular evolution, but also of the
formation of pseudobulges.   

Figure \ref{hd_hb} presents the correlation between disk and 
bulge scale lengths for a set of spirals. Filled circles correspond to
our data and open ones are those compiled by \citet{beijer1999}. A
Kolmogorov-Smirnov (KS) test show that the idea that the two variables
are not correlated is rejected at the 98\% level. This could suggest
that disk and bulge formation  were coupled \citep{beijer1999}. Note
that there is {\em no correlation} between the bulge-to-disk scale
lengths and the central surface brightness of the disk, as shown by
Figure \ref{ratio_scale_lenghts_and_mu}. In this Figure, it is
apparent that most scale lengths ratios are around 0.05 and 0.2, with
four outliers with $h_b/h_d \ge 0.3$. These correspond to galaxies
470, 447, 324 y 264. They do have larger bulges, and well defined
surface brightness profiles, except galaxy 264 (with no acceptable
bulge fitting in the near-IR), and do not present any other different
feature from the rest of the sample, except galaxy 470, which present
inverse color gradient in its bulge. 

\placefigure{hd_hb}
\placefigure{ratio_scale_lenghts_and_mu}

The correlation between scale lengths of disks and bulges shown in Figure 
\ref{hd_hb} suggests that the formation of these two components is
coupled, supporting the hypothesis that disks form first and bulges
emerge after and, the larger the disk, the larger the corresponding
bulge scale length. The restricted range of $\log h \sim 0.1$ for the
bulge and disk scale lengths ratio (Figures \ref{mu_h} and 
\ref{ratio_scale_lenghts_and_mu}) is the same as
used as an argument for secular evolution models
\citep{courteau1996}: self consistent numerical simulations of disk
galaxies evolve toward a double exponential profile with a typical
bulge-to-disk ratio $\sim 0.1$ \citep{friedli1995}, suggesting that
bulge formation is  triggered by disk instability. If 
this scenario is correct, then bulges should be relatively young. More
recently, \citet{mayer2004} presented numerical simulations showing
that bars can be formed in massive LSBs under some specific constraint
of disk density and gas temperature. These bars, short in length
compared to the halo and quite unstable in time, {\em necessarily}
trigger the formation of a bulge component, similar to that present
in many LSBs observed in red and near-IR bands by \citet{oneil2000a}
and \citet{galaz2002}. 

For our sample of 21 galaxies (Figure \ref{hd_hb}, filled circles), we
define the relative size of 
their bulges using the ratio between their scale lengths ($h_b/h_d$),
which is naturally correlated with the Hubble type. A small bulge
has a ratio $h_b/h_d$ below 1$\sigma$  
compared to average (a Hubble type typically a Sc). A large bulge
has a $h_b/h_d$ above 1$\sigma$ respect to the average (a Hubble type
Sa or Sb). Although we do not observe a tight relationship
between these two scale lengths (particularly if we consider the 3
outliers marked in the figure), the trend is clear and similar to what
has been observed by \citet{beijer1999} (open circles), suggesting a
possible secular evolution. 
%Inside the error bar and the observed dispersion, we do 
%not observe a conspicuous correlation between these two scale lengths
%(suggesting a possible secular evolution). 
%Also, we do not 
%observe any indication that allow us to conclude that bulges and disk of
%LSBs evolve differently from those hosted by HSB galaxies: 
Both, $h_b$
and $h_d$, follow approximately the same trend, except that our slope
($<h_d/h_b> = 8.18$, $\sigma = 5.5$) is somewhat different to that of
\citet{beijer1999}, indicated as a dashed line in Figure \ref{hd_hb}
($<h_d/h_b> = 9.98$, $\sigma = 3.78$). We therefore define a bulge as
``small'' if $h_d/h_b > 13.8$ and ``large'' if $h_d/h_b < 2.68$,
considering our average slope $<h_d/h_b>$ and its measured standard
deviation. Note that we are not excluding any galaxy from our sample
following this criterion and, although the number of
galaxies is small, trends are observed clearly for the scale length
ratios. In Figure \ref{hb_hd_near_ir} we show a similar correlation as
in Figure \ref{hd_hb}, but for the $J$ and $K_s$ near-IR bands. The
correlation holds and in fact, the scatter is even smaller. However,
near-IR bulge scale lengths are larger compared to their optical
counterparts. This is natural if we think that the bulge is metal rich
compared to the disk and with older stellar populations. Therefore the
bulge appear brighter and larger in the near-IR. 

\placefigure{hb_hd_near_ir}

In summary, (1) the observed relationship between $h_b$, $h_d$ and
its ratios, (2) the inverse color gradients in the optical, (3)
the double exponential fit for most of the galaxies, both in the
optical and the near-IR bands, and (4) the almost null $J - K_s$ color
gradients, support the secular evolution hypothesis.

However, the fact that most bulges are redder than disks (see
Figure \ref{profiles}), favors at first impression the 
scenario where bulges were formed before disks. \citet{andreakis1998}
showed that, in this case, it should also exist a correlation between 
the bulge and disk scale lengths, similar to the correlation
observed in Figure  \ref{hd_hb}. If this last scenario is correct,
bulges should be relatively old compared to disks. Nevertheless, note
that even two stellar populations formed at the same time, but with
different surface density (e.g. the bulge and the disk, the first a
denser one) could evolve differently. A larger surface density would
favor a more efficient stellar formation, and therefore a larger SFR
at the bulge, resulting in a redder bulge compared to the disk.  
It is worth noting that the observed trend does not strongly support the
hypothesis that HSB galaxies are progenitors of LSB galaxies, since
we do not observe that HSB galaxies are bluer than LSB galaxies. In
fact, bulges of LSB galaxies appear metal poor compared to bulges of
HSB galaxies (see Figure \ref{BR_JK_2}).

One of the most striking features in the optical color
profiles, is the inverse color gradient in the central zone
in 11 of the 21 galaxies, i.e. the color is bluer as radius decreases.
\citet{boissier2003}, comparing data of LSB galaxies 
with chemical and spectrophotometric
models of galaxy evolution, arrive to the conclusion that, given the 
large scatter observed in the LSB galaxy properties, it is necessary
to introduce both starburst events as well as interruption of
ongoing stellar formation processes, in order to match models with 
observations. These events could arise by gravitational interactions,
causing then inverse color gradients. \citet{menanteau2001}, studying
a HST dataset of 77 early-type galaxies presenting this inverse color
gradient, use a multizone single-collapse model which account for
the observed blue cores. The model adopts a broad spread in formation
redshifts for elliptical galaxies, allowing some of these galaxies to
begin their formation up to 1 Gyr before the redshift of observation. 
Therefore, the single-zone collapse model produces cores that are
bluer than the outer regions because of the increase of the local
potential well toward the center, which makes star formation more
extended in the central region of the galaxy than in the outer
parts\footnote{Note that a recent by \citet{menanteau2004} present
evidence arguing that elliptical blue cores are due to the presence of
an AGN.}. 
Could these inverse color
gradients be an evidence of secular evolution? The answer to this question is
directly related to the age-metallicity degeneracy. Indeed, bulges can 
be redder because they are older, more metallic or more affected by
dust reddening. The inverse color gradients presented here would support
secular evolution, if it would be possible to show that these
gradients are exclusively due to metallicity
\citep{galaz2005}. However, further evidence already discussed here, namely 
(1) the relationship between $h_b$, $h_d$ and
its ratios, (2) the double exponential fit for most of the face-on
nucleated galaxies,
both in the optical and the near-IR bands (pseudobulges?), and (4) the almost null
radial $J - K_s$ color gradients, support the secular evolution
hypothesis. The null $J - K_s$ gradient suggests that the bulge and
disk stellar population are much more similar, i.e., that their
metallicity is uniform (see Figure \ref{BR_JK_1} or \ref{BR_JK_2}).

\section{Conclusions}

We have presented an analysis of a sample of 21 face-on nucleated HSB
and LSB spirals selected from the catalogue of \citet{impey1996}. We
have studied their bulges colors and structural parameters in scope of
bulge and disk formation, obtaining the following relevant results. 

First, bulge and disk scale lengths appear to be marginally
correlated, weakly supporting a secular evolutionary process
between these two components. However, the relationship between the
bulge and disk sizes for HSB seems (B/D $\sim 0.1$) to be identical to
that observed for LSB galaxies, which can be understood in a general
context of bulge and disk formation. Also, our results agree
with the analysis both from observations presented by
\citet{courteau1996}, and from models by \citet{mayer2004}, suggesting
that bulges could be formed by bar instabilities in early stages of
disk evolution. The recently discovered red LSBs, with a notable
non-axisymmetric structure and bulge components, are in agreement with
many of the LSB models which incorporate massive disks. 

Second, structural analysis shows that LSB galaxies follow the same 
trends (bulge size, disk size, scale lengths, etc.) observed in HSB
galaxies, which constitutes evidence that,  
at least in their structure, LSB galaxies are not
a different type of galaxy. On the other hand, we rule out the possibility that
HSB galaxies are progenitors of LSBs, basically using the fact that
the colors of their bulges are redder compared to those
observed in LSBs.

Third, almost half of our sample shows inverse gradients in their $B - R$
radial color profiles (bluer color toward center). Although there is not a
straightforward explanation for this, this analyze is key to understand 
correctly the possible chronology of events which lead to the
formation of disk and bulge. Our near-IR surface brightness profiles,
do not show inverse color gradients, but present 
null color gradients which, with the evidence that almost all the
light profiles are well fitted by exponential forms (in the optical
and in the near-IR), support the secular evolution for these face-on
nucleated galaxies and the existence of pseudobulges. 

Fourth, using a spectrophotometric model of galaxy evolution, and 
using the fact that the color index $J - K_s$ is metal sensitive, we
find evidence suggesting that bulges of LSBs are metal poor compared
to those hosted by HSBs, extending the overall result from emission
line analysis that stellar formation regions of LSBs are also metal
poor compared to  HSB ones \citep{deblok1998, galaz2005}. 
Using the synthesis model PEGASE, we 
have shown that a possible small amount of dust in the bulges does not play an
important role on the bulges color and structural parameter
properties, and that the extinction in $B - R$ is smaller than 0.3
mags. We arrive to the same conclusions of \citet{bell2000}. Moreover, we obtain
that bulges hosted by LSB galaxies tend to be small ones. In turn,
small bulges are metal poor, compared to larger bulges. This agree with the secular
evolution model, in which bulges are built up secularly from the disk,
where small bulges have been subject of weak stellar formation
processes and slower accretion, leading to a metal poor stellar
population in average.

\acknowledgements

This research has made use of the NASA/IPAC Extragalactic Database
(NED) which is operated by the Jet Propulsion Laboratory, California
Institute of  Technology, under contract with the National Aeronautics
and Space Administration (NASA). It also made use of the
VizieR catalogue access tool, CDS, Strasbourg, France. GG acknowledges
the support of FONDECYT project \#1040359. GG and LI thank FONDAP \#
15010003 ``Center for Astrophysics''. The authors thank the Las
Campanas Observatory staff for their help during the
observations. Finally, it is a pleasure to thank the referee, Dr. Karen
O'Neil, who made valuable corrections, suggestions and comments,
allowing to improve the quality of the paper.

\newpage
%
% Figure Captions
%
\begin{center}
\underline{Figure captions}% (Figures are not included)} 
\end{center}
% Figure 1
\figcaption[f1.eps]{Mosaic showing $B$ images of the 21 selected
spirals for this study. Note that all of them present bulges and are
face-on, except galaxy 515, for which we have estimated an inclination
of 45 deg, and therefore increasing the surface brightness, due to
inclination $i$, by $2.5\log(b/a)$, where $b=acos(i)$, and $a$ the major
axis. \label{mosaics}} 
% Figure 2
\figcaption[f2a.eps,f2b.eps,f2c.eps,f2d.eps,f2e.eps,f2f.eps,f2g.eps,f2h.eps,f2i.eps,f2j.eps,f2k.eps]{Optical
surface brigthness and color profiles for all 21 galaxies selected for the
analysis. Top panels for each galaxy represent the surface brightness
profile in the $B$ and $R$ band. We include the central surface
brightness of both the disk and bulge component, from the linear fit
represented by the dashed lines. The bottom panel represent the $B -
R$ color gradient. Vertical lines mark the approximate boundaries
obtained visually of the pure bulge and disk components. Dotted lines
indicate the central $r^{1/4}$ fit. \label{profiles}}  
% Figure 3
\figcaption[f3a.eps,f3b.eps,f3c.eps,f3d.eps,f3e.eps,f3f.eps]{Near-IR
surface brightness profiles and color gradients for the 21 face-on
nucleated galaxies. Most of the galaxies are well 
fitted by a combination of two exponetial profiles (disk and bulge),
except galaxies 59, 100, 224, 470, and 471, which bulge is better
fitted by a $r^{1/4}$ profile. Upper curves show the $K_s$ profile and
lower curves the $J$ profile. For clarity, dashed lines show the bulge and disk
exponential fits only for one of the two bands. Solid lines show the corresponding
bulge + disk fit. The vertical line indicate the maximum error
for the computed surface brightness and color, for each
panel. \label{near_ir_profiles}}  
% Figure 4
\figcaption[f4.eps]{$B$ absolute magnitude as a function of the $B
- R$ color for spirals investigated in this paper. Open circles denote
the 21 face-on spirals and filled circles denote other spirals and irregular
galaxies. The error bar in the bottom right side of the figure 
indicate the maximum error in absolute magnitude and
color. \label{color_mag_BR}}  
% Figure 5
\figcaption[f5.eps]{Disk central surface brightness in the $B$ band
as a function of the disk scale length. Black dots represent our
sample and open circles galaxies from
\citet{dejong1995}. Note the outlier LSB 264, with a disk scale length
of 0.34. This is a galaxy with a dominant bulge respect the disk (see
Figure 1). \label{mu_h}} 
% Figure 6
\figcaption[f6.eps]{Bulge central surface brightness in the $B$
band as a function of the disk central surface brightness in the $B$
band. Symbols as in Figure \ref{mu_h}. \label{mu_mu}}
% Figure 7
\figcaption[f7.eps]{Inner 2 kpc $B - R$ color as a function of the
disk central surface brightness in the $B$ band. Symbols as in Figure
\ref{mu_h}. \label{BR_mu}} 
% Figure 8
\figcaption[f8.eps]{Figures presenting the relationship between
the HI gas mass and various measured parameters. (a) as a function of
the color $B - R$ for the inner 2 kpc of each galaxy. Redder galaxies
tend to be massive in HI; (b) and (c) as a function of the total
absolute magnitudes in the $B$ and $R$ bands, respectively. Brighter
galaxies are also more massive in HI; (d) as a function of the $R$
surface brightness. Black dots represent
the 21 nucleated, face-on galaxies. Galaxies with low surface
brightnesses tend to have small HI masses, as
expected. \label{relations}}  
% Figure 9
\figcaption[f9.eps]{$J - K_s$ vs.  
$B - R$ color-color diagram for the bulge inner 2
kpc of our selected galaxies (black dots), and for those HSB galaxies
compiled by \citet{peletier1996} (open circles). 
The grid corresponds to the colors predicted by GISSEL96
spectrophotometric model of galaxy evolution by
\citet{bruzual2003}. Horizontal lines represent different
metallicities and vertical lines different ages since burst (see text
for detail). \label{BR_JK_2}}
%Figure 10
\figcaption[f10.eps]{$J - K_s$ vs.  
$B - R$ color-color diagramme for the bulge inner 2
kpc, for small bulges (open symbols) and for large bulges (filled
symbols). The grid as in Figure \ref{BR_JK_2}. 
It is apparent that small bulges appear younger and less metallic
compared to large ones. Small bulges tend to be metal poor compared to
large bulges. Also note that LSB galaxies (squares) tend to have small
bulges. Colors are corrected for Galactic extinction using the maps
of \citet{schlegel1998} and for internal extinction (see text for
details) \label{BR_JK_1}} 
% Figure 11
\figcaption[f11.eps]{$B - R_c$ (black points and solid
line) and $J - K_s$ (open squares and dashed line) color difference
for an evolving bulge when 60\% more surface density of dust is
considered initially. The fiducial dust content is taken as the dust surface
density of the Galaxy bulge. Each point is the color and color
difference at a given age. Numbers indicate age after bulge
formation. From one evolutionary stage to the following,
metallicity and dust surface density is
changing according to stellar evolution and winds. \label{pegase_reddening}}
% Figure 12
\figcaption[f12.eps]{Disk scale length in the $B$ band ($h_d$,
in kpc) as a function of the bulge scale length in the same band
($h_b$), for the galaxies in our sample (black dots) 
and by \citet{beijer1999} (open circles). Lines represent the
$h_d/h_b$ average ratio for our sample (solid line) and such for the
sample of \citet{beijer1999} (dashed line). \label{hd_hb}}
%Figure 13
\figcaption[f13.eps]{Bulge to disk scale length ratio, as a
function of the disk central surface brightness ($B$ band). Solid
circles denote results for the face-on sample and open circles other
spirals and irregular galaxies.  \label{ratio_scale_lenghts_and_mu}}
% Figure 14
\figcaption[f14.eps]{Near-IR disk scale length ($h_d$ in kpc) as a function
of the bulge scale length ($h_b$ in kpc), for the galaxies in our
sample. Filters as indicated in each panel. \label{hb_hd_near_ir}}  

\newpage

\begin{deluxetable}{lccccccccccc}
\tabletypesize{\small} 
\renewcommand{\arraystretch}{0.6} 
\tablecaption{Solutions for the photometric calibration given by equations 
1 and 2.}
\tablewidth{18cm}
\tablecolumns{12}
\tablehead{
\colhead{Night} &
\colhead{$Z_B$\tablenotemark{a}} &
\colhead{$k_B$\tablenotemark{b}} &
\colhead{$C_B$\tablenotemark{c}} &
\colhead{$Z_R$\tablenotemark{a}} &
\colhead{$k_R$\tablenotemark{b}} &
\colhead{$C_R$\tablenotemark{c}}}
\startdata
28-29 May 2000 & -0.49(0.02) & -0.20(0.01) & 0.15(0.01) & -0.61(0.03) & -0.10(0.02) &  0.01(0.016) \\
29-30 May 2000 & -0.58(0.57) & -0.13(0.50) & 0.15(0.02) & -0.72(0.53) & -0.04(0.45) &  0.01(0.013) \\
30-31 May 2000 & -0.65(0.43) & -0.17(0.32) & 0.15(0.01) & -0.69(0.36) & -0.06(0.33) &  0.01(0.011) \\
25-26 Aug 2000 & -0.91(0.09) & -0.13(0.02) & 0.06(0.01) & -0.42(0.15) &  0.01(0.08) &  0.01(0.009) \\
26-27 Aug 2000 & -0.89(0.14) & -0.16(0.10) & 0.07(0.01) & -0.53(0.14) &  0.01(0.11) &  0.02(0.012) \\
07-08 Apr 2002 & -0.98(0.18) & -0.08(0.17) & 0.02(0.04) & -0.46(0.23) & -0.01(0.23) & -0.06(0.044) \\
08-09 Apr 2002 & -0.89(0.11) & -0.22(0.09) & 0.07(0.01) & -0.45(0.10) & -0.09(0.08) & -0.01(0.013) 
\enddata
\tablenotetext{a}{Photometric zero point in the respective filter with its r.m.s.
	error in parenthesis.}
\tablenotetext{b}{Extinction coefficient in the respective filter with its r.m.s.
	error in parenthesis.}
\tablenotetext{c}{Color term coefficient in the respective filter with its r.m.s.
	error in parenthesis.}
\label{phot_coefs}
\end{deluxetable}

\newpage
%
% Tables
%
%
\begin{table}
\scriptsize
\renewcommand{\arraystretch}{0.6} 
\caption{$B$ and $R$ magnitudes and colors for all spirals in the
sample.}
\begin{tabular}{lllllllllll}
\hline\hline
No.& Name & $\mu_B($2kpc$)$ & $\mu_R($2kpc$)$ & $M_B$ & $M_R$ &
$(B-R)_C$ & $(B-R)_T$ & log(M$_{HI}$/M$_\odot$) & Type & D$_{eff}$\\
(1)&(2)&(3)&(4)&(5)&(6)&(7)&(8)&(9) & (10) & (11) \\ \hline
4 & 0013-0034 & 20.02(0.06) & 18.28(0.07)   &  -20.810.05) & -22.02(0.06) &
1.69(0.09) & 1.33(0.08) & 10.18 & Sc &  36.8 \\
16 & 0027+0134 & 22.22(0.06) & 21.13(0.07)  &  -17.20(0.06) & -18.10(0.07) &
1.03(0.09) & 0.85(0.08) & 8.65 & Sd & 18.8 \\
36 & 0104+0140 & 20.63(0.06) & 19.09(0.07)  &  -18.06(0.08) & -19.26(0.07) &
1.50(0.10) & 1.17(0.08) & 8.74 & Sb & 10.2 \\
{\bf 59} & 0121+0128 & 19.57(0.06) & 17.87(0.07)  &  -20.11(0.06) & -21.66(0.05) &
1.64(0.10) & 1.67(0.09) & 9.64 & Sc & 80.4 \\
{\bf 100} & 0233+0012 & 22.44(0.06) & 21.28(0.07) &  -17.19(0.05) & -18.16(0.05) &
1.11(0.09) & 1.09(0.08) & 8.89 & Sc & 46.8 \\
126 & 0311+0241 & 20.62(0.06) & 19.74(0.07) &  -19.23(0.06) & -20.09(0.08) &
0.73(0.09) & 0.89(0.08) & 9.57 & Irr &  12.2 \\
146 & 0336+0212 & 23.15(0.06) & 22.16(0.07) &  -16.24(0.08) & -17.07(0.07) &
0.80(0.09) & 0.65(0.08) & 8.62 & dIn & 24.4 \\
{\bf 196} & 0913+0054 & 20.04(0.01) & 18.65(0.02) &  -19.71(0.07) & -20.97(0.06) &
1.34(0.02) & 1.21(0.02) & 9.68 & Sm & 16.4 \\
200 & 0918-0028 & 21.25(0.01) & 20.23(0.01) &  -17.78(0.05) & -18.79(0.05) &
0.98(0.02) & 0.95(0.02) & 8.63 & Sb 16.2 \\
{\bf 207} & 0929+0147 & 20.77(0.10) & 18.57(0.13) &  -19.88(0.06) & -21.50(0.05) &
2.07(0.16) & 1.49(0.14) & 10.03 & Sc & 26.6 \\
{\bf 213} & 0954+020 & 20.63(0.10) & 19.62(0.13)  &  -19.48(0.04) & -20.43(0.07) &
0.97(0.16) & 0.90(0.14) & 9.59 & Sc & 18.6 \\
{\bf 224} & 1007+0121 & 20.94(0.01) & 18.80(0.02) &  -21.23(0.06) & -22.81(0.05) &
2.08(0.02) & 1.52(0.02) & 10.69 & Sc & 16.2 \\
225 & 1008+0128 & 21.05(0.01) & 19.57(0.02) &  -19.61(0.05) & -20.72(0.07) &
1.41(0.02) & 1.03(0.02) & 9.87 & Irr & 17.4 \\
{\bf 242} & 1030+0252 & 19.90(0.06) & 18.63(0.05) &  -19.87(0.08) & -21.04(0.07) &
1.22(0.08) & 1.12(0.07) & 9.65 & SB & 15.2 \\
253 & 1036+0158 & 22.47(0.01) & 21.58(0.01) &  -10.62(0.04) & -11.47(0.05) &
0.83(0.02) & 0.79(0.02) & 7.62 & dIn &  23.8 \\
{\bf 264} & 1043+0202 & 24.24(0.05) & 23.25(0.05) &  -10.71(0.06) & -11.69(0.07) &
0.91(0.07) & 0.90(0.07) & 7.73 & dIn & 13.6 \\
266 & 1047+0131 & 24.07(0.01) & 23.04(0.02) &  -12.68(0.06) & -13.60(0.05) &
0.95(0.02) & 0.84(0.02) & 7.94 & dI & 15.6 \\
270 & 1050+0253 & 22.81(0.01) & 22.24(0.01) &  -12.80(0.05) & -13.39(0.04) &
0.50(0.02) & 0.52(0.02) & 8.02 & dIn & 17.4 \\
295 & 1124-0043 & 22.39(0.01) & 21.62(0.02) &  -13.08(0.05) & -13.83(0.06) &
0.73(0.02) & 0.70(0.02) & 8.10 & dIn & 25.8 \\
308 & 1156+0254 & 23.91(0.01) & 22.92(0.01) &  -14.36(0.04) & -15.26(0.05) &
0.93(0.02) & 0.84(0.02) & 8.72 & Sc & 28.4 \\
{\bf 324} & 1211+0226 & 21.53(0.05) & 19.80(0.05) &  -20.22(0.07) & -21.40(0.08) &
1.69(0.07) & 1.14(0.07) & 10.19 & Sc & 29.8 \\
329 & 1216+0029 & 24.14(0.10) & 23.40(0.15) &  -10.67(0.12) & -11.30(0.17) &
0.70(0.19) & 0.59(0.16) & 7.55 & dIn & 21.6 \\
330 & 1217+0103 & 22.25(0.01) & 21.34(0.01) &  -16.13(0.04) & -16.89(0.05) &
0.88(0.02) & 0.73(0.02) & 8.62 & Sm & 34.2 \\
{\bf 345} & 1226+0105 & 20.38(0.01) & 19.23(0.02) &  -19.68(0.05) & -20.82(0.05) &
1.11(0.02) & 1.10(0.02) & 10.26 & Sc &  25.8 \\
349 & 1228+0157 & 23.00(0.01) & 22.16(0.01) &  -12.97(0.04) & -13.82(0.04) &
0.81(0.02) & 0.82(0.02) & 7.75 & dIn & 23.8 \\
365 & 1257+0219 & 22.40(0.01) & 21.59(0.02) &  -11.55(0.05) & -12.53(0.07) &
0.76(0.02) & 0.94(0.02) & 7.61 & dIn & 26.2 \\
370 & 1300+0144 & 21.35(0.06) & 19.78(0.06) &  -19.86(0.08) & -21.12(0.07) &
1.53(0.08) & 1.22(0.07) & 9.98 & Sc &  15.0 \\
{\bf 377} & 1310-0019 & 22.44(0.05) & 21.11(0.05) &  -19.12(0.07) & -20.10(0.07) &
1.28(0.07) & 0.94(0.07) & 9.84 & Sc &  19.4 \\
378 & 1315+0029 & 21.92(0.01) & 20.87(0.02) &  -18.59(0.04) & -19.48(0.05) &
1.00(0.02) & 0.84(0.02) & 9.94 & Sm & 25.6 \\
380 & 1321+0137 & 20.79(0.01) & 19.06(0.02) &  -20.45(0.06) & -21.77(0.07) &
1.69(0.02) & 1.27(0.02) & 9.64 & Sm &  13.4 \\
384 & 1326+0109 & 23.84(0.09) & 22.70(0.13) &  -16.45(0.13) & -17.33(0.15) &
1.09(0.16) & 0.83(0.14) & 8.74 & dI & 35.8 \\
385 & 1327+0148 & 23.24(0.01) & 22.49(0.01) &  -12.49(0.05) & -13.24(0.06) &
0.70(0.02) & 0.71(0.02) & 7.49 & dIn & 21.4 \\
393 & 1350+0022 & 22.44(0.01) & 21.63(0.01) &  -17.62(0.04) & -18.35(0.04) &
0.74(0.02) & 0.67(0.02) & 8.84 & Sm & 8.6 \\
398 & 1353+020 & 23.51(0.09) & 22.88(0.13)  &  -15.91(0.10) & -16.49(0.16) &
0.58(0.16) & 0.52(0.14) & 8.96 & dI & 16.6 \\
400 & 1357-0017 & 23.89(0.05) & 22.89(0.05) &  -15.16(0.07) & -15.99(0.08) &
0.93(0.07) & 0.77(0.07) & 8.74 & Sm & 20.2 \\
407 & 1401+0108 & 21.27(0.01) & 19.90(0.02) &  -19.96(0.04) & -20.88(0.05) &
1.31(0.02) & 0.86(0.02) & 9.66 & Sb & 11.2 \\
{\bf 410} & 1405+0006 & 21.39(0.01) & 20.27(0.02) &  -18.99(0.05) & -19.85(0.06) &
1.06(0.02) & 0.79(0.02) & 9.73 & Sb & 12.4 \\
424 & 1433+0249 & 24.12(0.01) & 23.06(0.02) &  -13.04(0.05) & -14.02(0.07) &
1.01(0.02) & 0.92(0.02) & 8.03 & dI & 30.2 \\
433 & 1438+0049 & 23.32(0.10) & 21.06(0.15) &  -15.14(0.14) & -16.15(0.18) &
2.18(0.19) & 0.94(0.16) & 8.30 & dIn & 25.0 \\
435 & 1439+0053 & 22.96(0.01) & 21.73(0.02) &  -15.71(0.04) & -16.72(0.06) &
1.15(0.02) & 0.94(0.02) & 8.53 & Sd & 30.4 \\
437 & 1440-0008 & 23.35(0.01) & 22.33(0.02) &  -14.03(0.05) & -14.95(0.07) &
0.96(0.02) & 0.85(0.02) & 7.92 & dIn & 20.8 \\
446 & 1446+0231 & 23.19(0.01) & 22.20(0.02) &  -17.08(0.05) & -17.89(0.07) &
0.91(0.02) & 0.73(0.02) & 9.63 & Sm & 13.4 \\ 
{\bf 447} & 1446+0238 & 21.47(0.01) & 19.89(0.02) &  -19.33(0.04) & -20.54(0.06) &
1.51(0.02) & 1.15(0.02) & 9.73 & Sc & 19.0 \\ 
462 & 2303-0006 & 21.20(0.07) & 19.86(0.09) &  -20.28(0.09) & -21.21(0.11) &
1.27(0.12) & 0.86(0.10) & 9.71 & Sm & 27.4 \\ 
{\bf 463} & 2304+0155 & 21.72(0.06) & 20.63(0.07) &  -18.63(0.08) & -19.57(0.09) &
0.99(0.10) & 0.84(0.09) & 9.57 & SB & 27.2 \\ 
{\bf 468} & 2311-000 & 20.94(0.07) & 19.76(0.08)  &  -18.87(0.09) & -19.80(0.12) &
1.10(0.10) & 0.86(0.09) & 8.95 & Sa & 16.4 \\ 
{\bf 470} & 2312-0011 & 20.71(0.06) & 18.94(0.07) &  -20.33(0.08) & -21.63(0.09) &
1.70(0.09) & 1.23(0.08) & 9.98 & Sb & 98 11 \\ 
{\bf 471} & 2313+0008 & 20.92(0.07) & 19.57(0.09) &  -20.19(0.08) & -21.29(0.12) &
1.27(0.12) & 1.18(0.10) & 9.76 & Sb & 21.2 \\ 
{\bf 473} & 2315-0000 & 21.51(0.07) & 19.44(0.08) &  -19.87(0.09) & -21.24(0.11) &
2.00(0.10) & 1.47(0.09) & 9.71 & Sc & 24.6 \\ 
474 & 2317+0112 & 20.83(0.06) & 19.40(0.07) &  -20.19(0.08) & -21.29(0.09) &
1.35(0.10) & 1.20(0.09) & 9.74 & Sb & 20.2 \\ 
{\bf 484} & 2320+0110 & 19.57(0.06) & 18.60(0.07) &  -20.27(0.08) & -21.09(0.09) &
0.91(0.09) & 0.93(0.08) & 9.91 & Sb & 13.0 \\ 
485 & 2320+0107 & 20.63(0.06) & 19.51(0.07) &  -19.48(0.08) & -20.46(0.10) &
1.05(0.09) & 1.09(0.08) & 9.84 & Sb &  11.0 \\ 
{\bf 488} & 2327-0007 & 22.27(0.06) & 20.83(0.07) &  -18.52(0.08) & -19.76(0.09) &
1.36(0.09) & 1.16(0.08) & 8.99 & Sc & 38.8 \\ 
492 & 2329+0203 & 22.04(0.06) & 20.92(0.07) &  -17.77(0.09) & -18.76(0.10) &
1.04(0.09) & 0.91(0.08) & 8.99 & Sc &  18.4 \\ 
{\bf 515} & 2349+0248 & 21.05(0.06) & 19.77(0.07) &  -19.35(0.08) & -20.27(0.09) &
1.20(0.09) & 0.85(0.08) & 9.79 & Sbc & 31.2 \\ \hline
\end{tabular}
\\
Notes:   \\ 
(1) Correlative number to the \citet{impey1996} catalogue. Bold face
identifications denote the 21 face-on galaxies studied in detail in
this paper. \\
(2) Name from the \citet{impey1996} catalogue. \\
(3) Surface brightness of the inner 2 kpc diameter in the $B$ band. \\
(4) Surface brightness of the inner 2 kpc diameter in the $R$ band. \\
(5) Absolute magnitude in the $B$ band, to the isophote $\mu_B = 25.0$
\mucentral. \\ 
(6) Absolute magnitude in the $R$ band, to the isophote $\mu_R = 24.0$
\mucentral. \\ 
(7) $B-R$ color in the inner (2 kpc) galaxy region. \\ 
(8) Total $B-R$ color. \\
(9) HI mass in solar masses, taken from \citet{impey1996}. \\
(10) Hubble type from \citet{impey1996} catalogue. \\
(11) Effective diameter (in arcsec), the diameter at which half of the
light is included, from \citet{impey1996}.  
\label{magnitudes}
\end{table}

\begin{table}
\renewcommand{\arraystretch}{0.6} 
\caption{Optical structural parameters for the selected 21 face-on, nucleated
spiral galaxies.}
\begin{tabular}{l|llll|llll}
\hline\hline
\multicolumn{5}{c}{$B$} & \multicolumn{4}{c}{$R$} \\ \hline
No. & $\mu_{0,B}$ & $\mu_{0,D}$ & $h_{B}$ & $h_{D}$ & $\mu_{0,B}$ & 
$\mu_{0,D}$ & $h_{B}$ & $h_{D}$ \\
(1) & (2) & (3) & (4) & (5) &  (2) & (3) & (4) & (5) \\ \hline

59 & 18.31 & 21.89 & 0.53 & 10.4 & 17.04 & 20.19 & 0.72 & 8.45 \\ 	
100 & 20.62 & 22.49 & 0.19 & 4.27 & 19.40 & 21.42 & 0.19 & 3.95	\\
196 & 18.73 & 21.34 & 0.56 & 4.89 & 17.23 & 19.99 & 0.54 & 4.47	\\
207 & 19.95 & 21.63 & 1.01 & 6.52 & 16.81 & 20.30 & 0.94 & 9.85	\\
213 & 19.56 & 22.34 & 0.80 & 5.96 & 18.66 & 21.17 & 0.91 & 5.08	\\
224 & 20.42 & 22.82 & 2.18 & 25.1 & 18.25 & 20.90 & 1.96 & 18.5	\\
242 & 18.50 & 20.98 & 0.53 & 3.38 & 16.85 & 19.63 & 0.40 & 3.07	\\
264 & 22.05 & 22.91 & 0.15 & 0.34 & 20.84 & 22.02 & 0.14 & 0.34	\\
324 & 21.17 & 21.93 & 2.58 & 6.54 & 19.33 & 20.58 & 1.93 & 5.76	\\
345 & 19.60 & 22.55 & 1.20 & 11.7 & 18.56 & 21.81 & 1.36 & 12.6	\\
377 & 22.09 & $\dots$&	3.60 & $\dots$&	20.72 & $\dots$& 3.01 &
$\dots$ \\
410 & 20.74 & $\dots$&	1.55 & $\dots$&	19.70 & $\dots$& 1.67 &
$\dots$ \\
447 & 20.78 & 21.74 & 1.24 & 4.09 & 19.23 & 20.54 & 1.34 & 4.10	\\
463 & 20.97 & $\dots$&	1.85 & $\dots$&	19.94 & $\dots$& 1.92 &
$\dots$ \\
468 & 20.20 & $\dots$&	0.92 & $\dots$&	19.11 & $\dots$& 1.14 &
$\dots$ \\
470 & 20.09 & 20.52 & 1.34 & 4.04 & 18.46 & 19.72 & 1.44 & 4.71	\\
471 & 19.43 & 20.72 & 0.57 & 4.20 & 18.39 & 19.67 & 0.74 & 4.21	\\
473 & 19.16 & 21.42 & 0.90 & 4.99 & 17.73 & 20.04 & 0.99 & 4.70	\\
484 & 18.40 & $\dots$&	0.70 & $\dots$&	17.58 & $\dots$& 0.79 &
$\dots$ \\
488 & 21.14 & 22.65 & 0.62 & 5.02 & 19.67 & 21.46 & 0.59 & 4.95	\\
515 & 20.18 & 21.78 & 0.87 & 5.50 & 18.89 & 20.71 & 0.87 & 4.96	\\ \hline
\end{tabular}
\\
(1) ID number correlated with the \citet{impey1996} catalogue. \\
(2) Bulge central surface brightness in units of mag arcsec$^{-2}$. \\
(3) Disk central surface brightness in units of mag arcsec$^{-2}$. \\
(4) Exponential scale length of the bulge in units of kpc. \\
(5) Exponential scale length of the disk in units of kpc. \\
{\bf Note}: Dots indicate that no disk has been detected with a signal
to noise ratio larger than 1.5.  
\label{structural_parameters}
\end{table}

\begin{table}
\renewcommand{\arraystretch}{0.6} 
\caption{Near-IR structural parameters for the selected 21 face-on,
nucleated spirals selected.} 
\begin{tabular}{ll|llllll|llllll}
\hline\hline
\multicolumn{8}{c}{$J$} & \multicolumn{6}{c}{$K_s$} \\ \hline
No. & $cz$ & $\mu_{0,B}$ & $h_{B}$ & $\mu_{0,D}$ & $h_{D}$ & $\mu_{eff}$ & r$_{eff}$ 
& $\mu_{0,B}$ & $h_{B}$ & $\mu_{0,D}$ & $h_{D}$ & $\mu_{eff}$ & r$_{eff}$ \\
(1) & (2) & (3) & (4) & (5) &(6) & (7) & (8) & (3) & (4) & (5) & (6) &
(7) & (8) \\ \hline
  59 & 5028.0 & $\dots$  &$\dots$  & 22.33 &  40.29 & 22.33 &  20.62 & 
 $\dots$ & $\dots$  & 21.21 & 33.96 & 21.08  & 14.55 \\ 

  100 & 2615.0  & $\dots$  & $\dots$ & 24.74 &  15.66 & 28.76 & 40.38 & $\dots$
&$\dots$  & 24.02 & 13.88 & 27.75 &  24.73 \\

  196 & 11401.0 &  15.57 &  0.23   & 21.70 &  2.93  & $\dots$  &$\dots$
   & 14.90
  & 0.23 & 20.64 &  3.07 & $\dots$  & $\dots$  \\
  207 & 17324.0 &  17.80 &  2.15   & 22.04 &  29.68 & $\dots$  & $\dots$ & 16.70
  &  2.16 & 20.36 & 17.17 & $\dots$ & $\dots$ \\
  213 & 9592.0  &  14.72 &  0.31   & 19.16 &  3.16 & $\dots$ & $\dots$ & 16.55
  &  0.71 & 18.77 &  5.09 & $\dots$ &  $\dots$ \\

  224 & 29213.0 & $\dots$  &$\dots$   & 18.87 &  12.60 &  18.92 &  3.12 & $\dots$
 &  $\dots$  & 17.46 & 10.98 & 16.37 &  1.62 \\

  242 &  8825.0 &  18.43 &  1.53   & 20.90 &  13.27 & $\dots$ & $\dots$ & 17.58
  &  1.60  & 20.06 & 13.34 & $\dots$ & $\dots$ \\
  264 & 1018.0  & $\dots$   & $\dots$    & 22.36 &  0.82  &$\dots$ &$\dots$ & $\dots$
  & $\dots$  & 21.55 &  0.82 &$\dots$ & $\dots$\\
  324 & 22259.0 &  22.00 &  5.40   & 22.50 &  22.07 &$\dots$ &$\dots$ & 20.50
  &  7.71 & 21.70 & 28.19 &$\dots$ & $\dots$ \\
  345 & 23655.0 &  19.50 &  6.49   & 22.62 &  43.36 &$\dots$ &$\dots$ & 18.72
  &  6.30 & 21.95 & 46.28 &$\dots$ & $\dots$ \\
  377 & 12040.0 &  22.56 &  7.79   & 23.82 &  37.69 &$\dots$ &$\dots$ & 22.26
  &  5.68 & 23.05 & 36.22 & $\dots$&$\dots$ \\
  410 & 7518.0  &  22.37 &  1.92   & 21.61 &  10.04 &$\dots$ &$\dots$ & 21.77
  &  1.88 & 20.96 &  9.38 &$\dots$ &$\dots$ \\
  447 & 10281.0 &  21.23 &  2.36  & 21.14 &  11.60 &$\dots$ &$\dots$ & 20.44
  &  7.63 & 22.02 & 23.46 &$\dots$ & $\dots$\\
  463 & 5244.0  &  21.78 &  3.06   & 22.54 &  17.22 & $\dots$&$\dots$ & 21.14
  &  2.53 & 21.76 & 17.63 &$\dots$ &$\dots$ \\
  468 & 4392.0  &  21.26 &  0.65   & 20.87 &  7.10  &$\dots$ &$\dots$ & 20.39
  &  1.00 & 20.29 &  7.43 &$\dots$ &$\dots$ \\

  470 & 15380.0 &$\dots$ & $\dots$ & 22.40 & 18.53 & 23.75 & 28.32 & $\dots$
 &$\dots$ & 21.07 &  5.37 & 24.11 & 75.59 \\

  471 & 8667.0  & $\dots$ & $\dots$ & 21.27 &  12.93 & 25.41 & 41.83 & $\dots$
& $\dots$ & 20.46 & 12.54 & 24.84 & 51.72 \\

  473 & 8938.0  &   $\dots$ &   $\dots$  &$\dots$   & $\dots$   &$\dots$ & $\dots$& 
 $\dots$ &$\dots$   & $\dots$  & $\dots$ &$\dots$ &$\dots$ \\
  484 & 8951.0  &  20.18 &  3.11   & 20.46 &  8.70  &$\dots$ &$\dots$ & 19.31
  &  3.03 & 19.89 &  9.33 &$\dots$ & $\dots$ \\
  488 & 5207.0  &  21.32 &  1.68  & 22.92 &  18.82 &$\dots$ &$\dots$ & 20.69
  &  1.86 & 22.34 & 23.58 &$\dots$ &$\dots$ \\
  515 & 5324.0  &  20.43 &  3.03   & 22.04 &  14.47 &$\dots$ &$\dots$ & 19.65
  &  2.74 & 21.18 & 12.56 &$\dots$ &$\dots$ \\ \hline
\end{tabular}
\\
(1) ID number correlated with the \citet{impey1996} catalogue. \\
(2) Heliocentric velocity, from \citet{impey1996} catalogue. \\
(3) Bulge central surface brightness in \mucentral. \\
(4) Exponential scale length of the bulge in kpc. \\
(5) Disk central surface brightness in \mucentral. \\
(6) Exponential scale length of the disk in kpc. \\
(7) Effective surface brightness in \mucentral. \\
(8) Effective radius in kpc. \\
\label{structural_parameters_ir}
\end{table}

\clearpage
\end{document}